\documentclass[12st]{article}

\usepackage{epsfig}

\begin{document}

\begin{center}
{\Large \bf Dynamical evolution of the Universe in the quark-hadron
phase transition and possible nugget formation}

\vskip .5cm
{\bf Deepak Chandra}\\ 
{\em  Physics Department, S. G. T. B. Khalsa College,}\\
{\em  University of Delhi, Delhi-110007, India.}\\
{\bf Ashok Goyal \footnote{E--mail:agoyal@ducos.ernet.in}}
\\
{\em  Department of Physics and Astrophysics,}\\
{\em  University of Delhi, Delhi-110007, India.}\\
{\em  and Inter University Center for Astronomy and Astrophysics,}\\
{\em  Ganeshkhind, Pune-411007, India.}\\

\end{center}

\begin{abstract}
\noindent  We study the dynamics of first-order phase transition in the 
early Universe when it was $10-50 \mu s$ old with quarks and gluons condensing 
into hadrons. We look at how the Universe evolved through the phase 
transition in small as well as large super cooling scenario, specifically 
exploring the formation of quark nuggets and their possible survival.
The nucleation of the hadron phase introduces new distance scales in 
the Universe, which we estimate along with the hadron fraction, temperature,
nucleation time etc. It is of interest to explore whether there is a relic
signature of this transition in the form of quark nuggets which might
be identified with the recently observed dark objects in our galactic
halo and account for the Dark Matter in the Universe at present.
\end{abstract}
\pagebreak

\indent It is well known that a phase transition from quark gluon plasma to 
confined hadronic matter must have occurred at some point in the evolution of 
the early Universe, typically at around $10-50 \mu s$ after the Big Bang. This
leads to an exciting possibility of the formation of quark nuggets through 
the cosmic separation of phases {\cite{wit}}. As the temperature of the 
Universe falls below the the critical temperature $T_c$ of the phase
transition, the quark gluon plasma super cools and the transition proceeds
through the bubble nucleation of the hadron phase. The typical
distance between the nucleated bubbles introduces a new distance scale to the
Universe which depends critically on the super cooling that takes place. 
This bubble nucleation of the hadron phase is governed by the formation of
critical bubbles, that is bubbles bigger than the critical size start to
expand and coalesce till the entire Universe is filled with the hadron phase.
As the hadronic bubbles expand, they heat the surrounding plasma, shutting off
further nucleation and the two phases coexist in thermal equilibrium. The
 hadron phase expands driving the deconfined quark phase into small regions of
space and it may happen that the process stops after the quarks reach 
sufficiently high density to provide enough pressure to balance the surface
tension and the pressure of the hadron phase. The quark matter trapped in 
these regions constitute the quark nuggets. The number of particles trapped in 
the quark nugget, its size and formation time are dependent sensitively on the
degree of super cooling. The duration of the phase transition also depends on 
the expansion of the Universe and on other parameters like the bag pressure $B$
and the surface tension $\sigma$. 
\par
The quark nuggets formed in the small super cooling scenario are in a hot
environment around the critical temperature $T_c$ and are susceptible
to evaporation from the surface {\cite{Alcock} } and to boiling
through subsequent  hadronic bubble nucleation inside the nuggets 
{\cite{Olinto}}. However in the large super cooling scenario we
have the interesting possibility of these nuggets forming at
a much lower temperature than $T_c$ due to the long time of phase
transition and consequent expansion of the Universe. Alcock and Farhi 
{\cite{Alcock}} have shown that the quark nuggets with baryon numbers 
$\leq 10^{52}$ and mass $\leq 10^{-5} M_{\odot}$ are unlikely to survive 
evaporation of hadrons from the surface. Boiling was shown to be even
more efficient mechanism of nugget destruction by Alcock and Olinto 
{\cite{Olinto}}. If these nuggets are formed at around $100 MeV$, they
cannot have more baryons than are allowed in the horizon by 
the standard model of Cosmology, i.e. $\sim 10^{49}(\frac{T}{100 MeV})^{-3}$ 
and would not survive till the present epoch. We will see that for
small super cooling this is the likely fate of the nuggets. These
results were somewhat modified by Madsen et.al. {\cite{Madsen}} by
taking into account the flavor equilibrium near the nugget surface
for the case of evaporation and by considering the effect of
interactions in the hadronic gas in a relativistic mean field model 
described by Walecka for the case of boiling {\cite{Mad1}}. They 
suggested that the quark nuggets with large baryon number allowed by the 
causality limit may after all be able to survive from the early Universe. In
the large super cooling scenario the time at which these nuggets are formed 
can be quite late along with a much lower formation temperature due to the
expansion of the Universe. Such nuggets can have a baryon number content of 
$\geq 10^{52}$ and being at a much lower temperature than $T_c$
$(\leq 0.1T_c)$ will easily survive till the present epoch. The
number of baryons in the horizon (of size $\sim 2t)$ can also be large due to
long time it takes for the bubbles to meet and form the nuggets.     
\par
There have been recent observations by gravitational micro lensing 
{\cite{Alcock1}} of dark objects in our galactic halo having masses of
about $0.01-1$ solar mass. If these objects have to be identified with
quark nuggets, they could only have been formed at a time later than
the time when the Universe cooled through $T_c$, i.e. later than
50-100 $\mu$s after the Big Bang. Such a possibility of the nuggets 
forming at a temperature $\sim 0.1 MeV$, implying a high degree of
super cooling and strongly first order phase transition was recently
investigated by Cottingham et. al. {\cite{cot}} in the Lee-Wick model
 {\cite{lw}}. Their investigation showed that the time of formation
and the baryon content of these nuggets are essentially determined by
the rate at which the hadron bubbles nucleate. However there was still
the question of reheating due to the expansion of the bubbles of
hadron phase which raises the temperature towards $T_c$. These studies
have also been carried out by taking interactions into account 
in both the phases and by incorporating the effects of curvature
{\cite{ag}} energy in the calculations.
\par
Studies of quark-hadron phase transitions in the early Universe 
{\cite{Fuller}}, in heavy ion collision {\cite{Csernai}} and in high density 
nuclear matter have been done previously by looking in detail at the 
dynamics of the phase transition. Kapusta et. al. have applied a 
recently computed nucleation rate {\cite{Csernai}} to
a first order phase transition in a set of rate equations to study the time 
evolution of a quark-gluon plasma as it converts to hadronic matter in heavy
ion collisions. Based on Bjorken hydrodynamics and on current parameter values,
they find the transition generates 30 percent extra entropy and also a time
delay of $\sim 11 fm/c$ in completion of the transition. Kajantie and
Kurki-Suonio \cite{Fuller} studied how the early Universe during the 
quark-hadron phase transition evolved through the mixed phase in a 
scenario with small initial super cooling. Fuller et.al.\cite{Fuller} have also 
studied the dynamics of the Universe during the constant temperature
 coexistence epoch.
\par
In this letter we calculate the nucleation rate and quantitatively study what 
happens when the temperature drops to $T_c$. This nucleation rate is used to 
solve a set of rate equations to study the time evolution of the quark-gluon 
phase as it converts to hadronic matter in an expanding Universe. A novel 
feature of this method is that reheating of the plasma during phase transition
is included in the calculations and all relevant quantities can be evaluated 
as a function of time.

\indent Bubble Nucleation:- When the early Universe as a quark-gluon 
thermodynamic system cools
through the critical temperature $T_c$, energetically, the new phase 
remains unfavorable as there is 
free energy associated with surface of separation between the phases. Small
volumes of new phase are thus unfavorable and all nucleated bubbles with
radii less than critical radius collapse and die out. But those with radii 
greater than the critical radius expand until they coalesce with each other. 
So super cooling occurs before the new phase actually appears and is then 
followed by reheating due to release of latent heat. 
The bubble nucleation rate {\cite{Fuller}} at temperature T is given by
\begin{equation}
I=I_o e^{-\frac{W_c}{T}} 
\end{equation}
where $I_o$ is the prefactor having dimension of $T^4$. The prefactor used 
traditionally in early Universe studies {\cite{Fuller}} is given by 
$I_o=(\frac{W_c}{2\pi T})^{\frac{3}{2}}T^4$. Csernai and Kapusta
{\cite{Csernai}} have recently computed this
prefactor in a coarse-grained effective field theory
approximation to QCD and give 
$I_o = \frac{16}{3\pi}
 (\frac{\sigma}{3T})^{\frac{3}{2}}\frac{\sigma \eta_q R_c}
{\xi^4_q (\Delta w)^2}$ 
 where $ \eta_q=14.4T^3 $  is the shear
 viscosity in the plasma phase, $\xi_q$ is a correlation length of 
 order 0.7 fm in the plasma phase and $ \Delta w$ is the difference
 in the enthalpy densities of the two phases. In this letter we use both
 these prefactors for comparison. The critical bubble radius $R_c$ and  
 the critical free energy $W_c$ are obtained by maximizing the thermodynamic
 work expended to create a bubble and are given by 
$R_c= \frac{2 \sigma}{P_h(T)-P_q(T)} $ and  
$W_c = 4\pi \sigma R^2_c (T)/3  = \frac{16\pi \sigma^3}{3\Delta P^2}$ 
where $\Delta P =P_h(T)-P_q(T)$  is the pressure difference in hadron
 and quark phase.
\par
For simplicity we describe the quark matter by a plasma of massless
$u$,$d$ quarks and massless gluons without interaction. The long range
non-perturbative effects are parameterized by the bag constant $B$. The
light particles (photons, neutrinos, and electrons) contribute equally
to the pressure in both the phases. The 
pressure in the QGP phase is given by $P_q(T)=\frac {1} {3} g_q \frac {\pi^2}
 {30} T^4-B$ where $g_q\sim 51.25$ is the effective number of degrees
of freedom. In the hadronic phase the pressure is
given by $P_h(T) =\frac{1}{3} g_h\frac{\pi^2}{30}T^4$ where $g_h\sim 17.25$,
taking the three pions as massless.
\par
The fraction of the Universe $h(t)$ which has been converted from QCD
plasma phase to hadronic phase at the time $t$ was first given by Guth
and Weinberg \cite{Guth} and applied to cosmological first order
phase transitions. Csernai and Kapusta \cite{Csernai} gave a kinetic
equation for calculating $h(t)$ which takes bubble growth into
account. If the early Universe cools to $T_c$ at time $t_c$, then at
same later time $t$ the fraction of the Universe in hadronic phase is
given by the kinetic equation
\begin{equation}
h(t)=\int_{t_c}^{t} I(T(t'))[1-h(t')]V(t',t)\bigg[\frac {R(t')} {R(t)}\bigg] ^3 dt'
\end{equation}
where $V(t',t)$ is the volume of a bubble at time $t$ which was
nucleated at an earlier time $t'$ and $R(t)$ is the scale factor.
This takes bubble growth into account
and can be given simply as
 \begin{equation}
V(t',t)=\frac{4\pi}{3}\Bigg[ R_c(T(t'))+\int_{t'}^{t} \frac{R(t)}
{R(t'')} v(T(t''))dt''\Bigg]^3
\end{equation}
where $v(T)$ is the speed of the growing bubble wall and can be taken
to be $v(T)=v_o [ 1- \frac{T}{T_c} ]^{\frac{3}{2}}$ where
$v_o=3c$. This has the correct behavior in that closer $T$ is to
$T_c$ slower do the bubbles grow. When $T=\frac{2}{3}T_c$ we have
$v(T)=\frac{1}{\sqrt{3}}$ the speed of sound of a 
massless gas. For $T<\frac{2}{3}T_c$ which occurs when there is
large super cooling, we use the value
$v(T)=\frac{1}{\sqrt{3}}$. In this analysis collision and
fusion of bubbles have not been taken into consideration. This seems
to be justified as far as fusion of bubbles is concerned. 
Witten \cite{wit} and Kurki Suonio \cite{Fuller} have shown
that for small enough bubbles, surface tension will cause them to
coalesce into larger bubbles. This distance scale has been estimated
 to be given by
$l_c=3(\frac{\sigma}{T_c^3})^{\frac{1}{3}}(\frac{T_c}{200 MeV})^
{-\frac{5}{3}}$ mm.
If nucleation scale $l_n$ is larger than $l_c$ then we are justified
in neglecting fusion of bubbles. As we will see, this is the case in 
all interesting scenarios.
\par
The other equation we need is the dynamical equation which couples time
evolution of temperature to the hadron fraction $h(t)$. We use the two
Einstein's equations as applied to the early Universe neglecting curvature.
\begin{equation}
\frac{\dot{R}}{R}=\sqrt{\frac{8\pi G} {3}} \rho^{\frac{1}{2}}
\end{equation}
\begin{equation}
\frac{\dot{R}}{R}=-\frac{1}{3w} \frac{d\rho}{dt} 
\end{equation}
where $w=\rho+P$ is the enthalpy density of the Universe at time $t$.
The energy density in the mixed
phase is given by $\rho(T)=h(t)\rho_h(T)+[1-h(t)]\rho_q(T)$, where
$\rho_h$ and $\rho_q$ are the energy densities in the two phases at
temperature $T$ and similarly for enthalpy. We have numerically
integrated the coupled dynamical equations (2),(4) and(5) to study the
evolution of the phase transition starting above $T_c$ at some
temperature $T$ corresponding to time $t$. This time $t$ has been
obtained by integrating the Einstein's equation (4) and (5) and is
given by 
\begin{equation}
t=-\sqrt {\frac {9}{24\pi G}} \int_{\infty}^{T}\frac {dT} 
{(\frac{g_q\pi^2T^6}{30}+BT^2)^{1/2}}
\end{equation}
The number density of nucleated sites at time $t$, is given by
\begin{equation}
N(t)=\int_{t_c}^{t} I(t')[1-h(t')]\bigg(\frac {R(t')} {R(t)}\bigg)^{3}dt'
\end{equation}
Therefore the typical separation between nucleation sites is 
 $l_n=N(t)^{-1/3}$.
This distance scale will eventually determine the number of quarks in
a nugget. This scale can be up to $10^{12}$ Km depending on the
parameters $B$ and $\sigma$ which
correspond to a distance of $\sim 1.4 Mpc$ today. Thus the structures
generated in this transition can be of the order of distance between
galaxies today. The observable separation of galaxies in the Universe
can be a remnant of this transition with the centers of the galaxies
being the quark nuggets. Of course the collision of bubbles and their
random nucleation and interaction will also lead to clustering of the
nuggets, which can qualitatively explain the clustering of galaxies.
\par
To get an idea about the super cooling before nucleation begins, we can
plot the nucleation time as a function of temperature, defined by 
$\tau^{-1}_{nucleation}=\frac{4\pi R_c^3}{3}I$. This characteristic
time scale neglects bubble growth. The quark
number density in the early Universe is given by 
$n_q=\frac{2}{\pi^2}\zeta (3)(\frac{n_q}{n_\gamma})T^3$ where $
\frac{n_q}{n_\gamma}$ is the quark to photon ratio estimated from the
abundance of luminous matter in the Universe to be roughly equal to 
$3\times 10^{-10}$. The quark nuggets which
have trapped the quarks in the plasma phase have $N_q$ quarks at time
$t$ given by the number of quarks in a volume 
$\frac{1}{N(t)}$, i.e. $N_q=\frac{n_q}{N(t)}$. The nucleation sites
are actually randomly distributed, but we expect a distribution of
quark numbers around $N_q$. The average temperature at which nuggets 
are formed when bubbles
coalesce is obtained by finding the average time at which the expanding bubble
surfaces meet. Assuming a cubic lattice , this is given by
setting twice the radius of the expanding bubbles equal to the length
scale $l_n$ of the lattice spacing. We have done this numerically to
get the corresponding time $t_f$ and temperature $T_f$ and these quantities
have been estimated for different values of $B$ and
$\sigma $. When the fraction of the space occupied by the bubbles is
around $50$ percent, we expect the bubbles to meet in an ideal
picture,i.e. if all bubbles are essentially nucleated at one instant
which is the maximum nucleation time and they all have the same
radius. However we have a distribution of expanding bubble sizes
because of the different points of time at which they were
nucleated. Therefore the estimate of the time of nugget formation by
treating all bubbles to be of the same size is an underestimate. We
find that hadron fraction $h(t)$ is only around $.12$ when bubbles
meet by this criteria. However we do not expect this to change
qualitatively the broad picture of the transition and the nugget
formation apart from reducing the formation time.
\par
In Fig.1 we have plotted the $log_{10}$ of nucleation time $\tau$ as a
function of temperature for different values of the bag pressure $B$,
surface tension $\sigma$ and the prefactor $I_o$. These curves clearly
show when nucleation becomes large and how much super cooling of the
Universe occurs. Fig.2 shows the temperature as a function of time. It
is clear from this diagram that reheating takes place as nucleation
begins . As $\sigma$ increases, the
super cooling is larger and the transition takes much longer to
complete with more chance of nugget formation. At the bottom of the
curves, nucleation starts with release of latent heat and consequent
rise in temperature. When the transition completes, the Universe again
starts cooling. However, larger the super cooling, slower is the
reheating as the Universe keeps on expanding. This allows the nuggets
to be formed at a much lower temperature when bubble walls meet. 
From Fig.3 we see that the
average bubble density $N(t)$ is initially zero and then increases
with time. As soon as reheating starts, bubble nucleation shuts off at
a particular point. The transition now continues only by expansion of
the nucleated bubbles. The fall in $N(t)$ beyond this point is due to
the expansion of the Universe. Fig.4 shows the fraction $h(t)$ of the
Universe in hadron phase as a function of time. For small values of
$\sigma$ the transition completes quickly as $h(t)$ goes to $1$. But
for larger $\sigma$ it takes a larger time for $h(t)$ to become
$1$. It may be comparable and even larger than the expansion time scale of
the Universe. Thus the dilution factor is very important here as it
makes the reheating much slower and consequently delays the
transition. We also notice that in the large super cooling scenario the Kapusta
prefactor becomes much bigger than the standard one by many orders of
magnitude. This makes the nucleation rate as well as the reheating
faster.In the case of low super cooling the two prefactors give 
almost identical results. The
number of quarks in the horizon $N_{qH}$ at time $t$ is 
$N_{qH}\sim n_q(\frac{4}{3}
\pi t^3 )\sim(\frac{n_q}{n_\gamma})\frac
{2\zeta(3)} {3\pi^2} T^34\pi t^3$
and we find that for all interesting cases $N_q \le N_{qH}$ and this number
is very sensitively dependent on the surface tension. Physically it is
possible to have $N_{qH}\ge N_q\ge 10^{52}$ for some values of the
parameters B and $\sigma$. In table I below we list some physical quantities 
for some representative values of $B$ and $\sigma$.

\[ {\begin{array}{cccccccc}
\hline
 B^{1/4} &\sigma & T_c & t_f & T_f & N_q & N_{qH} & l_n \\ 
 MeV  & MeV fm^{-2} & MeV & \mu s & MeV &   &  & m  \\
\hline
235 & 50 & 169 & 12.1 & 169 & 2.6\times 10^{28} & 7\times 10^{52}  & 8\times 10^{-3}	 \\
100 & 39.5 & 71.9 & 2595 & 13.9 & 1.2\times 10^{50} & 3.8\times 10^{56} & 8.4\times 10^{5}\\
113 & 57.1 & 81.3 & 5138 & 12.3 & 9\times 10^{49} & 2\times 10^{57} & 1.7\times 10^6\\	
\hline
\end{array}} \]  
Table I. Some relevant physical quantities for some representative values of 
$B$ and $\sigma$ using the standard prefactor.

\noindent As nuggets with $N_q\ge 10^{52}$ are expected to survive the
transition, they will contribute to the energy density of the
universe. However, nuggets of solar mass range need somewhat
unrealistically large surface tension, specially if the bag pressure
$B$ is also large.

\par
\indent In conclusion, detailed dynamics of the quark-hadron transition in the
early Universe show that the evolution of the Universe does not
necessarily follow the small super cooling scenario and certain choices
of B and $\sigma$ can have a bearing on the present state of the
Universe. We have explored in detail the possibility of nugget
formation and also estimated their average separation, time of
formation, quark content and survivability by solving a rate equation
coupled to the Einstein's equations in an expanding universe. Clearly,
the analysis can be improved by taking interactions into account in
both the phases and also bubble interactions may be incorporated in
the calculations. This will be reported elsewhere. however we believe
that qualitatively the results given here will hold. If the nuggets
studied above are indeed formed in a much cooler environment, they
could contribute significantly to the missing mass in the Universe and
be candidates for dark matter. 

\par
Acknowledgement: We would like to thank J. V. Narlikar for providing
hospitality at the Inter University center for Astronomy and
Astrophysics, Pune, India where a part of this work was completed.

\pagebreak


\begin{figure}
\epsfig{file=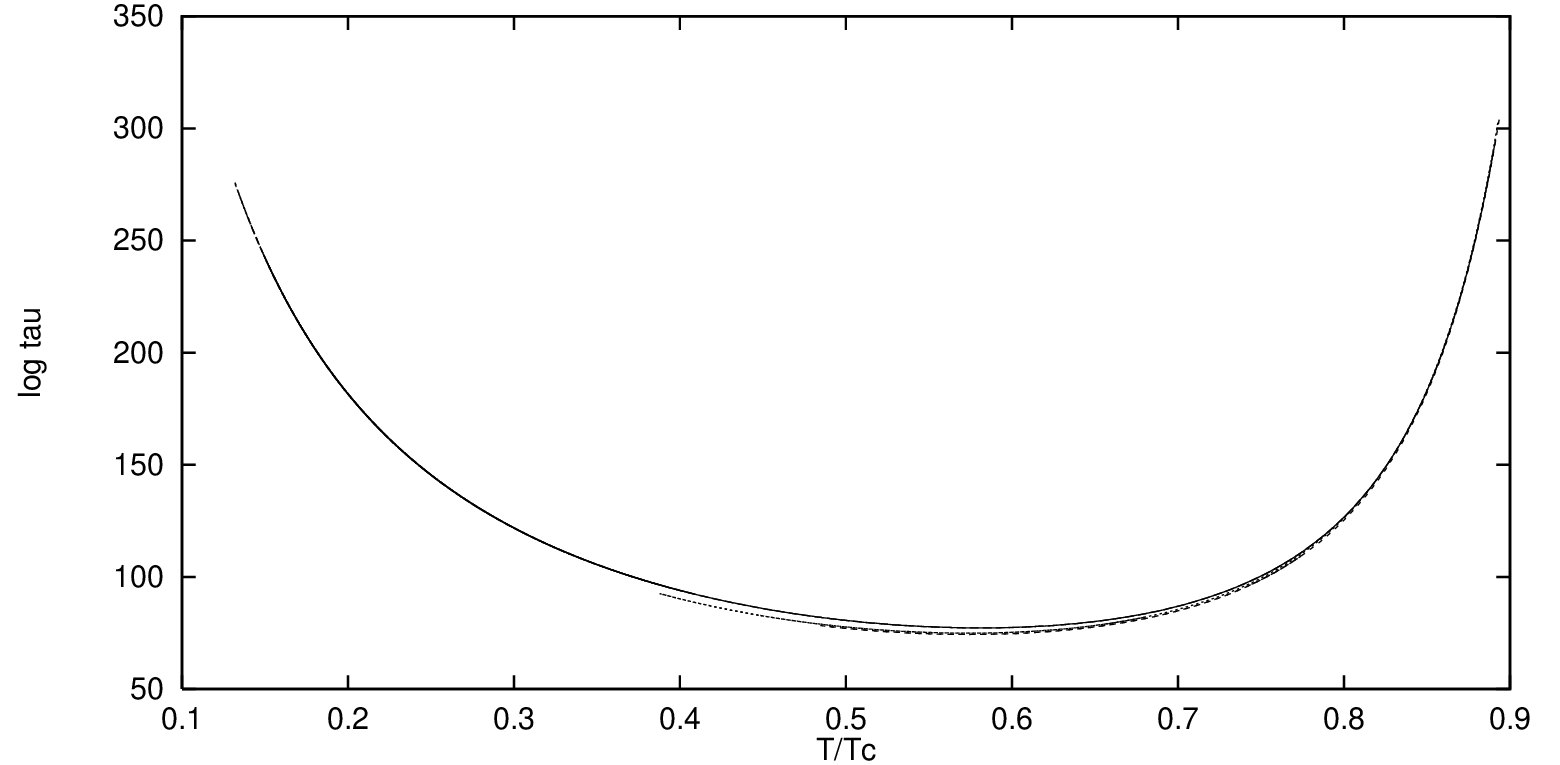,width=14cm ,height=10cm}
\caption{Log of the nucleation time $\tau$ in units of fm/c as a function of 
temperature. Solid and dashed curves are for $B^{1/4} =100 MeV$ and $\sigma =
39.5 MeV fm^{-2}$ with the standard and the Kapusta prefactors respectively.
Long dashed and dotted curves are for $B^{1/4} =113 MeV$ and $\sigma =
57.1 MeV fm^{-2}$ with the standard and the Kapusta prefactors respectively.}
\end{figure}
\begin{figure}
\epsfig{file=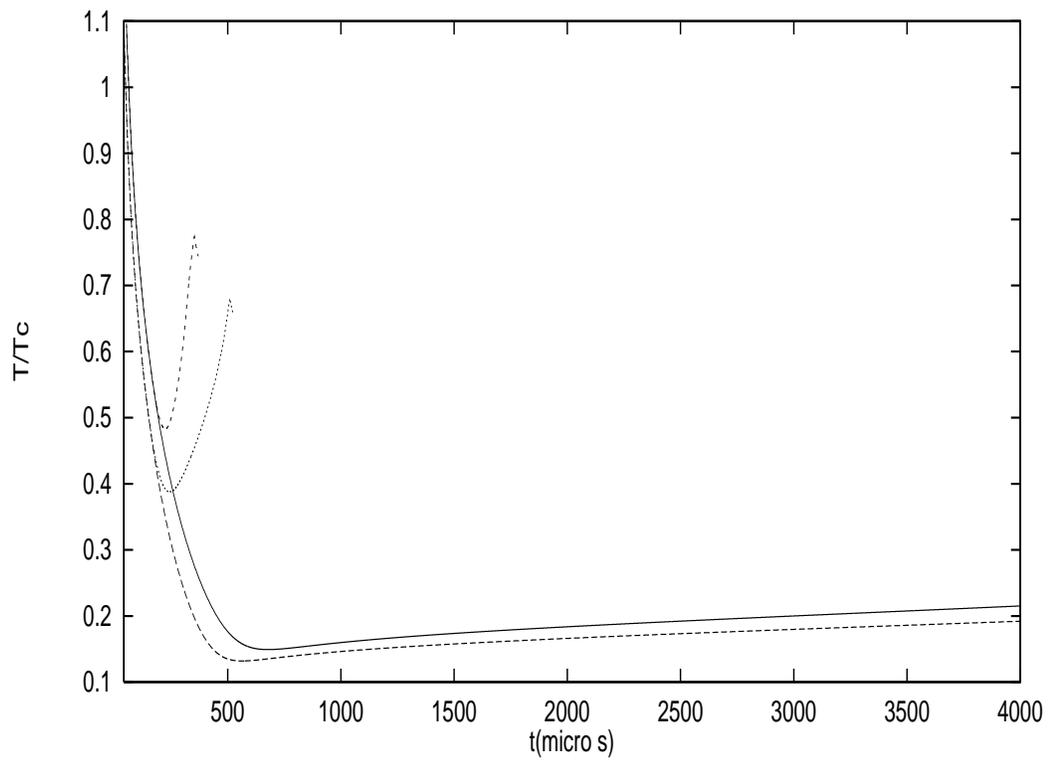,width=14cm,height=10cm}
\caption{Temperature as a function of time. Solid, dotted, dashed and long 
dashed curves are as in fig. 1.}
\end{figure}
\begin{figure}
\epsfig{file=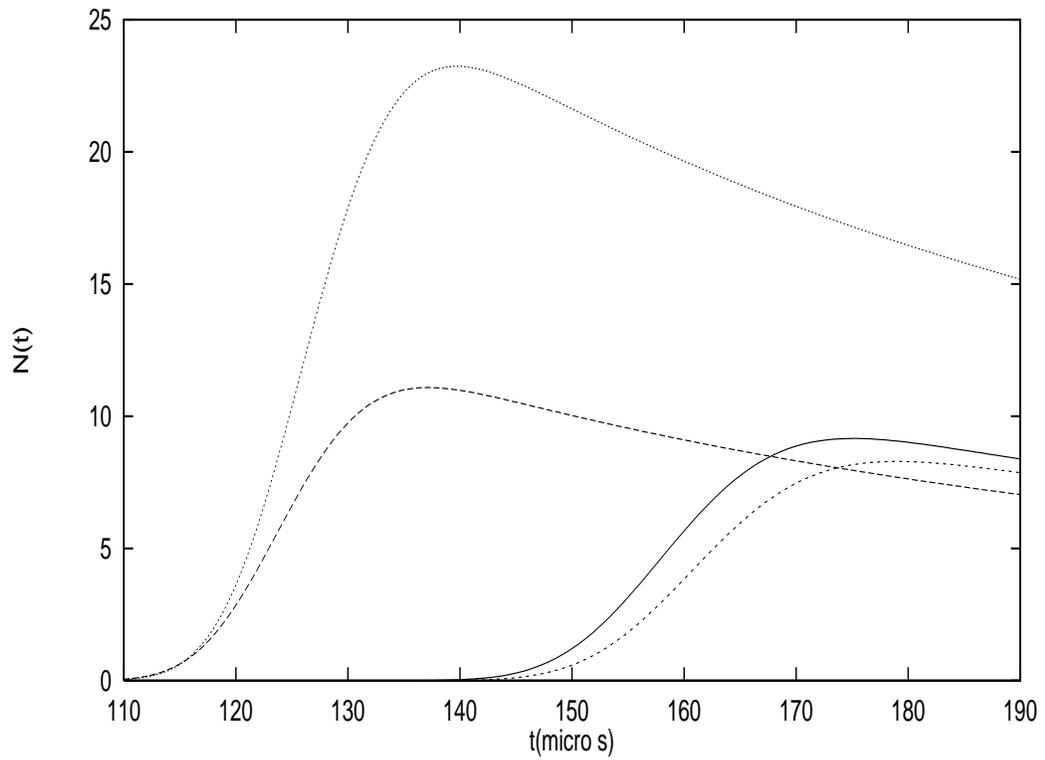,width=14cm,height=10cm}
\caption{Average bubble density $N(t)$ as a function of time in units of 
$(10^5 km^3)^{-1}$. The bubble density for the standard prefactor (solid 
and long dashed curves) is normalized by multiplying with a factor of $10^3$. 
Curves as in fig. 1.}
\end{figure}
\begin{figure}
\epsfig{file=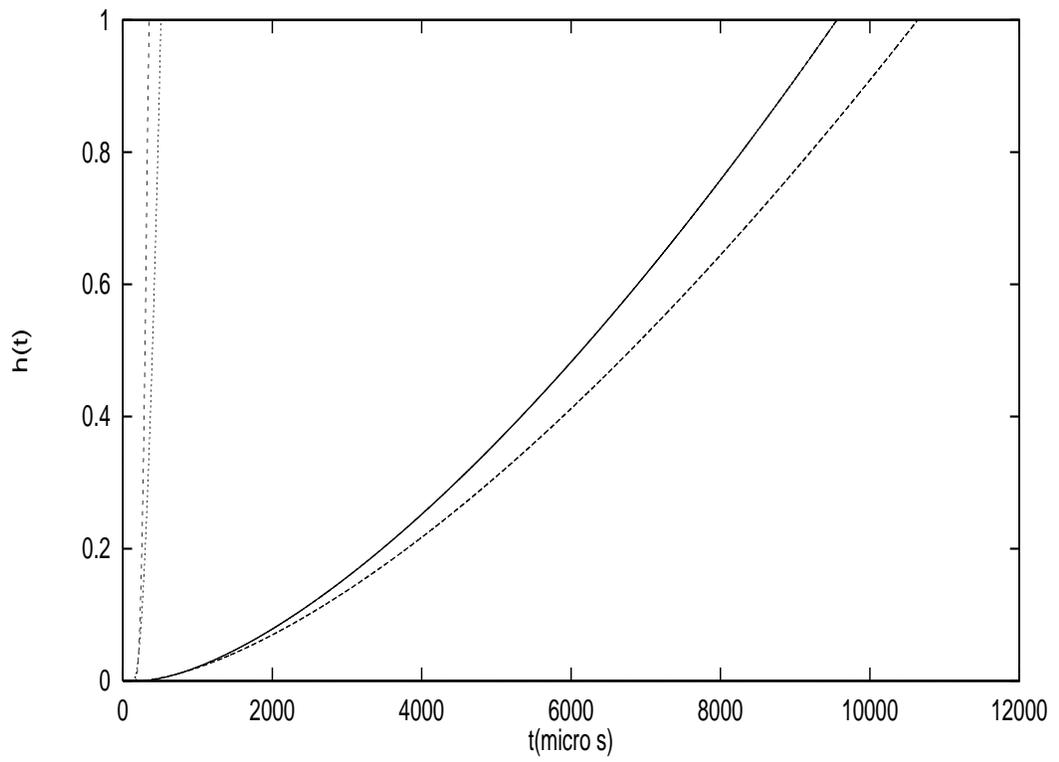,width=14cm,height=10cm}
\caption{The hadron fraction as a function of time. Curves as in fig. 1.}
\end{figure}

\end{document}